\documentclass[5p,times]{elsarticle}
\usepackage{lipsum}
\usepackage{lineno,hyperref}
\usepackage[pdftex]{color}
\usepackage[font=footnotesize,labelfont=bf]{caption}
\usepackage[font=footnotesize,labelfont=bf]{subcaption}
\journal{Journal of \LaTeX\ Templates}

\usepackage{colortbl}
\usepackage{xcolor}

\begin{document}

\begin{frontmatter}

\title{Pattern formation and coarsening dynamics in apparent competition models}

\author[ect-ufrn,ua]{J. Menezes\corref{mycorrespondingauthor}}
\cortext[mycorrespondingauthor]{Corresponding author}
\ead{jmenezes@ect.ufrn.br}
\address[ect-ufrn]{Escola de Ci\^encias e Tecnologia, Universidade
Federal do Rio Grande do Norte\\ Caixa Postal 1524, 59072-970 Natal,
RN, Brazil}
\address[ua]{Institute for Biodiversity and Ecosystem
Dynamics, University of Amsterdam, Science Park 904, 1098 XH
Amsterdam, The Netherlands}

\author[deb-ufrn,innels]{B. Moura} 
\ead{beatriz.moura@edu.isd.org.br}
\address[deb-ufrn]{Departamento de Engenharia Biomédica, Universidade Federal do Rio Grande do Norte\\
Av. Senador Salgado Filho, 300, 59078-970, Natal, RN, Brazil}
\address[innels]{Edmond and Lily Safra International Neuroscience Institute, Santos Dumont Institute\\
Av Santos Dumont, 1560, 59280-000, Macaiba, RN, Brazil}

\begin{abstract}
Apparent competition is an indirect interaction between species that share natural resources without any mutual aggression but negatively affect each other if there is a common enemy. The negative results of the apparent competition are reflected in the species spatial segregation, which impacts the dynamics of their populations.
Performing a series of stochastic simulations, we study a model where organisms of two prey species do not compete for space
but share a common predator. Our outcomes elucidate the central role played by the predator in the pattern formation and coarsening dynamics in apparent competition models.
Investigating the effects of predator mortality on the persistence of the species, we find a crossover between a curvature driven scaling regime and a coexistence scenario.
For low predator mortality, spatial domains mainly inhabited by one type of prey arise, surrounded by interfaces that mostly contain predators. We demonstrate that the dynamics of the interface network are curvature driven whose coarsening follows a scaling law common to other nonlinear systems. The effects of the apparent competition decrease for high predator mortality, allowing organisms of two prey species to share a more significant fraction of lattice. Finally, our results reveal that predation capacity in single-prey domains influences the scaling power law that characterises the coarsening dynamics. Our findings may be helpful to biologists to understand the pattern formation and dynamics of biodiversity in systems with apparent competition.
\end{abstract}


\end{frontmatter}


\section{Introduction}\label{Sec1}
There is plenty of evidence that the spatial segregation of species is crucial to the formation and stability of ecosystems \cite{ecology}. Spatial patterns result from a variety of interactions performed by individuals of different species \cite{nowak06evolutionaryDynamicsBOOK}. To comprehend how space influences species coexistence, researchers have made experimental and theoretical studies to observe the dynamics of biological systems \cite{Nature-bio}. A remarkable experiment with bacteria \textit{Escherichia coli} revealed a cyclic dominance among three strains \cite{Coli,bacteria}. However, the authors reported that the only way to maintain coexistence is whether the interactions are performed locally, creating spatial domains inhabited by individuals of the same species \cite{Allelopathy}. Similar results were observed in groups of lizards and coral reefs \cite{lizards,Extra1}.

Local interactions have also been verified to be fundamental to stabilise competition systems without cyclic dominance. For example, it has been reported a temporal and spatial variation of mortality of the two competing butterfly species \textit{Danaus plexippus} and \textit{D. chrysippus} \cite{But1}. Spatial segregation of the two specialist larval parasitoids \textit{Cotesia melitaearum} and \textit{Hyposoter horticola} attacking the butterfly \textit{Melitaea cinxia} were also observed\cite{But2}.
However, spatial segregation can also arise when species do not compete directly, i.e., there is no interference among their individuals or fight for resources \cite{Chesson}. This type of negative interaction, named apparent competition, occurs when species share a common threat \cite{Bonsall,Taylor}. The term \textit{apparent} was introduced by Holt in 1977 to represent a negative indirect interaction between species that share a natural enemy\cite{Holt}. The effects of the apparent competition in the spatial distribution of species appear in many biological systems like interactions among plants, birds, and mammals \cite{Australis,Boreal,Field}. One of the characteristics observed in apparent competition systems is a higher predation rate in patches with individuals of both types of prey \cite{Hyperpredation}. 

Despite the vast material of this class of species interactions in literature, mathematical and numerical models to describe population dynamics in apparent competition systems are scarce (see \cite{ApparentEquations,Hyperpredation,MathematicalImpulsive} for examples of analytical studies). This work aims to understand the dynamics of spatial patterns in systems of two prey species whose individuals can share spatial areas without any aggression or competition for space, serving as prey for a common predator. To this purpose, we use stochastic numerical simulations following an agent-based code widely used for investigating spatial interactions of biological systems both for cyclic dominance \cite{Reichenbach-N-448-1046,Avelino-PRE-86-036112,uneven,Moura,Anti1,anti2,neigh} and interference competition \cite{Avelino-PRE-86-031119,Pereira,PhysRevE.89.042710,PhysRevE.99.052310}. In our model: i) individuals of two prey species can share space without any direct interference or competition for space; ii) individuals of two prey species have the same chances of reproducing when empty space is available; iii) predators have the natality rate controlled by the local prey availability, with maximum predation capacity in patches with both types of prey; iv) predators' mortality has natural causes not resulting from spatial interaction; v) individuals' movement is random and happens with the same probability, irrespective of the species.

Our goal is to understand how local predator-prey interactions influence population dynamics and coexistence. We investigate the species segregation and theoretically predict the coarsening process leading the spatial patterns to undergo a scaling regime. Considering various scenarios for predation mortality and reproduction rate, we study the conditions for the predator persistence in patches where individuals of both types of species are available or consumption is limited to only one type of prey. 

The outline of this paper is as follows. In Sec.~\ref{Sec2}, we introduce the model and the stochastic simulations. In Sec.~\ref{Sec3}, we investigate the pattern formation and the dynamics of the spatial densities. In Sec.~\ref{Sec4}, a theoretical prediction for the scaling exponent of the coarsening dynamics is presented. The role of predator mortality and predation capacity on the dynamics of the spatial patterns are studied in Sec.~\ref{Sec5} and Sec.~\ref{Sec6}, respectively. Finally, our comments and conclusions appear in Sec.~\ref{Sec7}.
\begin{figure}[t]
\centering
\includegraphics[width=40mm]{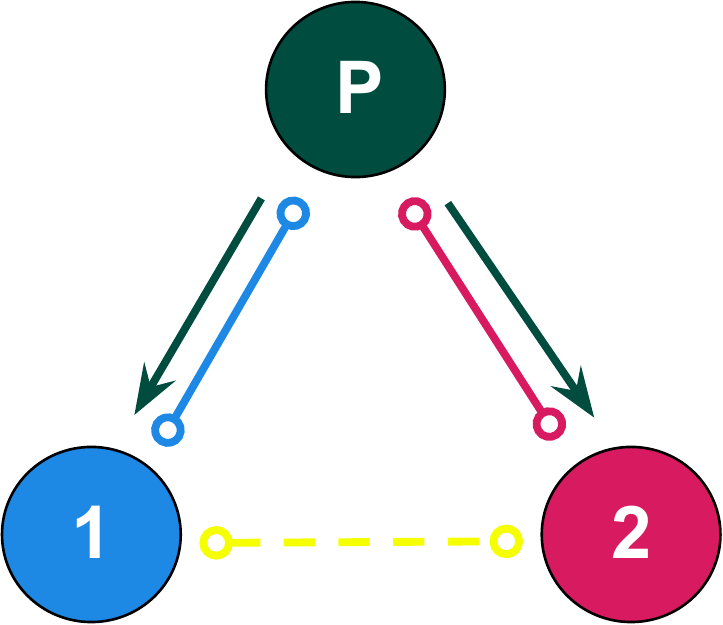}
\caption{Illustration of the predation and mobility rules in our apparent competition model.
Green arrows show predation interactions; blue and red lines indicate that prey and predator does not share the same space, while the dashed yellow line illustrates that species $1$ and $2$ do not compete for space.} 
\label{fig1}
\end{figure}
\begin{table}[h]
\caption{Set of interaction probabilities used in our stochastic simulations} 
\centering 
\begin{tabular}{c c} 
\hline\hline 
Interaction & Probability \\ [0.5ex] 
\hline 
predator mobility  & m \\
predation & p \\
predator mortality & d\\
prey mobility & m\\
prey reproduction & r\\[1ex] 
\hline 
\end{tabular}
\label{Table0} 
\end{table}
\begin{figure*}
\centering
 \begin{subfigure}{.19\textwidth}
        \centering
        \includegraphics[width=34mm]{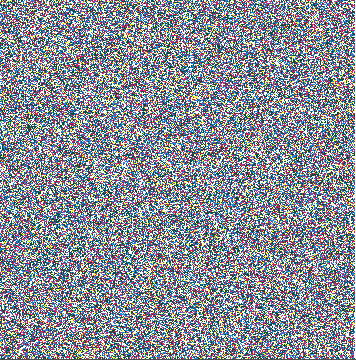}
        \caption{}\label{fig2a}
    \end{subfigure} %
    \begin{subfigure}{.19\textwidth}
        \centering
        \includegraphics[width=34mm]{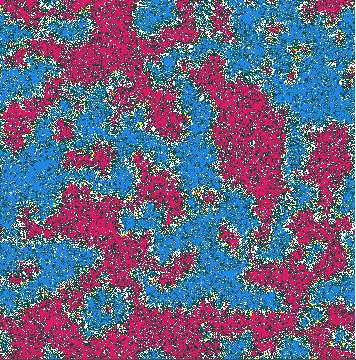}
        \caption{}\label{fig2b}
    \end{subfigure} %
       \begin{subfigure}{.19\textwidth}
        \centering
        \includegraphics[width=34mm]{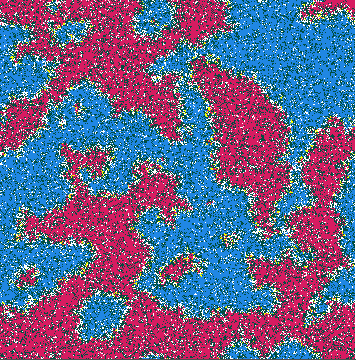}
        \caption{}\label{fig2c}
    \end{subfigure} %
   \begin{subfigure}{.19\textwidth}
        \centering
        \includegraphics[width=34mm]{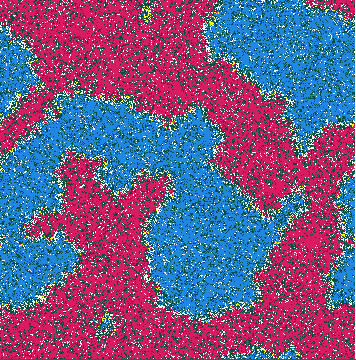}
        \caption{}\label{fig2d}
    \end{subfigure} %
    \begin{subfigure}{.19\textwidth}
        \centering
        \includegraphics[width=34mm]{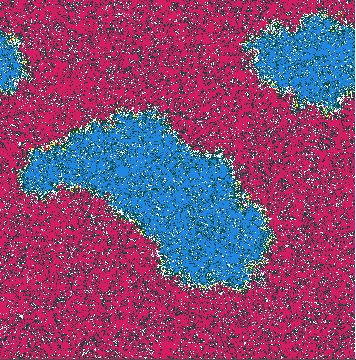}
        \caption{}\label{fig2e}
    \end{subfigure}\\
     \begin{subfigure}{.19\textwidth}
        \centering
        \includegraphics[width=34mm]{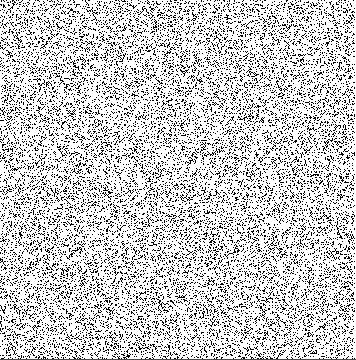}
        \caption{}\label{fig2f}
    \end{subfigure} %
       \begin{subfigure}{.19\textwidth}
        \centering
        \includegraphics[width=34mm]{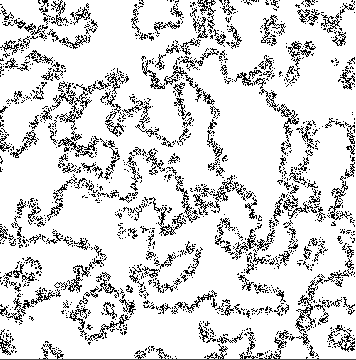}
        \caption{}\label{fig2g}
    \end{subfigure} %
   \begin{subfigure}{.19\textwidth}
        \centering
        \includegraphics[width=34mm]{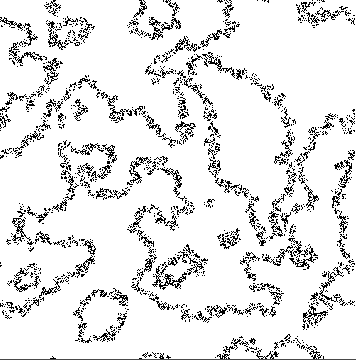}
        \caption{}\label{fig2h}
    \end{subfigure} %
   \begin{subfigure}{0.19\textwidth}
        \centering
        \includegraphics[width=34mm]{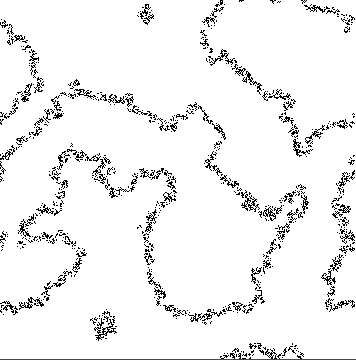}
        \caption{}\label{fig2i}
    \end{subfigure} %
   \begin{subfigure}{.19\textwidth}
        \centering
        \includegraphics[width=34mm]{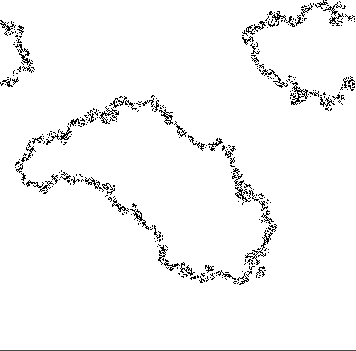}
        \caption{}\label{fig2j}
    \end{subfigure} %
\caption{Spatial patterns captured from a lattice with $500^2$ grid points running until $5000$ generations. Figures~\ref{fig2a} \ref{fig2b}, \ref{fig2c}, \ref{fig2d}, and \ref{fig2e} show the spatial distribution of organisms of prey species $1$ (blue regions), prey species $2$ (red areas) and predators (green patches) at the initial conditions and after $200$, $500$, $1000$ and $4000$ generations, respectively. White dots indicate empty space, while yellow indicates grid points shared by individuals of both types of prey. In Figs.~\ref{fig2f}, 
\ref{fig2g}, \ref{fig2h}, \ref{fig2i}, and \ref{fig2j}, black dots highlight the predators in patches with individuals of both prey species, whereas empty spaces, single-prey domains (indistinctly from the prey type), and predators consuming only one type of prey are left uncoloured. Videos https://youtu.be/9tDxXTkdePM and https://youtu.be/bdL2150lKfY show pattern formation and dynamics of the interface network during the whole simulation.}
  \label{fig2}
\end{figure*}


\section{The model} \label{Sec2}

We study an apparent competition model composed of one predator and two prey species, where individuals of different prey species do not compete for space. 
This means that two prey of different species can share the same spatial region without any interference competition. In contrast, regarding the occupation of space, the relationship between a predator and a prey of any species is anything but peaceful: besides predator-prey interactions, a prey and a predator compete for space, thus not living in the same spatial position. For this reason, in our stochastic simulations, each grid site $(x,y)$ may contain either a single prey, a pair of prey of distinct species, or a predator - otherwise, the grid point is an empty space. An individual prey is denoted by $i=1$ or $i=2$, where $i$ represents the species. 

The local availability of prey species defines the predator's performance: predation capacity is maximum whether organisms of both types of species are available for consumption. Therefore, each predator faces its reality,  consuming only the organisms of prey species present in its neighbourhood. Because of this, predation capacity in areas inhabited only by individuals of prey species $i$ is defined by the real parameter $\nu_i$, where $0 \leq \nu_{i} \leq 1$, with $i=1,2$; in patches with both types of prey, the predation capacity is $\nu_{1,2}$, where $0 \leq \nu_{1,2} \leq 1$. 
Throughout this paper, we assume that: i) predation capacity is maximum in patches where a predator can access individuals of both types of prey: $\nu_{1,2}\,=\,1$;
ii) predator's performance is the same in areas where organisms of only one type of prey are available, irrespective of the prey species: $\nu_1\,=\, \nu_2\,=\,\nu\,\leq\,1$.

At each time step one of these spatial interactions is implemented by our algorithm:
\begin{itemize}
\item
Prey reproduction: a prey of any species produces offspring at the grid point $(x,y)$, only if neither another individual of the same species nor a predator is present there;
\item
Prey mobility: a lonely prey changes positions with either another individual of the same species, a predator, or an empty space $(x,y)$;
\item
Predation: a predator consumes either a single individual or a pair of prey of distinct species present at the spatial position $(x,y)$, generating an offspring to fill that grid point.
\item
Predator mobility: a predator switches position with any content of the spatial position $(x,y)$.
\item
Predator mortality: a predator at $(x,y)$ dies naturally, leaving an empty space.
\end{itemize}
Figure \ref{fig1} illustrates the stochastic predation and mobility rules. Green arrows show that individuals of species $1$ and $2$ serve as prey for the predator; solid blue and red lines indicate that predators compete for space with species $1$ and $2$, respectively; yellow dashed lines illustrate that there is no competition for space between prey $1$ and prey $2$.
To implement the interactions, the algorithm follows the probabilities in Table~\ref{Table0}, with $m+p+d=1$ for a predator and $m+r =1$ for a prey.

We run simulations in square lattices of $\mathcal{N}$ sites, with periodic boundary conditions, where individuals of each species are positioned in different grid layers. We assum random initial conditions, where each grid site takes one of the possible configurations. Initially, the total numbers of individuals of every species are the same. The simulation algorithm follows three steps: i) selecting a random occupied grid point to be the active position; ii) drawing one of its eight neighbour sites to be the passive position (Moore neighbourhood); iii) randomly choosing an interaction to be executed by one individual at the active position - in case of two prey occupy the active position, the code randomly chooses one of them to be the active individual. If the active and the passive individuals match the raffled interaction, the implementation is realised and one timestep is counted; otherwise, the code repeats the three steps. Our time unit is called generation, which is the necessary time to $\mathcal{N}$ interactions to occur.

To determine the local predation capacity, we verify the presence of organisms of prey species $1$ and $2$ in the vicinity of a predator located at $(x,y)$. Defining a frequency neighbourhood as a disc of radius $R$ centred at $(x,y)$, we assume that predation capacity reaches its maximum value whether there are individuals of both types of prey within the predator's frequency neighbourhood. 
To quantify the population dynamics, we compute the species densities, defined as the fraction of the grid occupied by organisms of the species at time $t$. We denote the spatial species densities as $\rho_i$,  with $i\,=\,1,\,2,\,3$, where $i=1$  indicates prey species $1$, $i=2$ refers to prey species $2$, and $i=3$ represents all predators. In addition, we compute the predator density in patches with both types of prey, denoted by $\rho_4$.
\begin{figure}
\centering
\includegraphics[width=87mm]{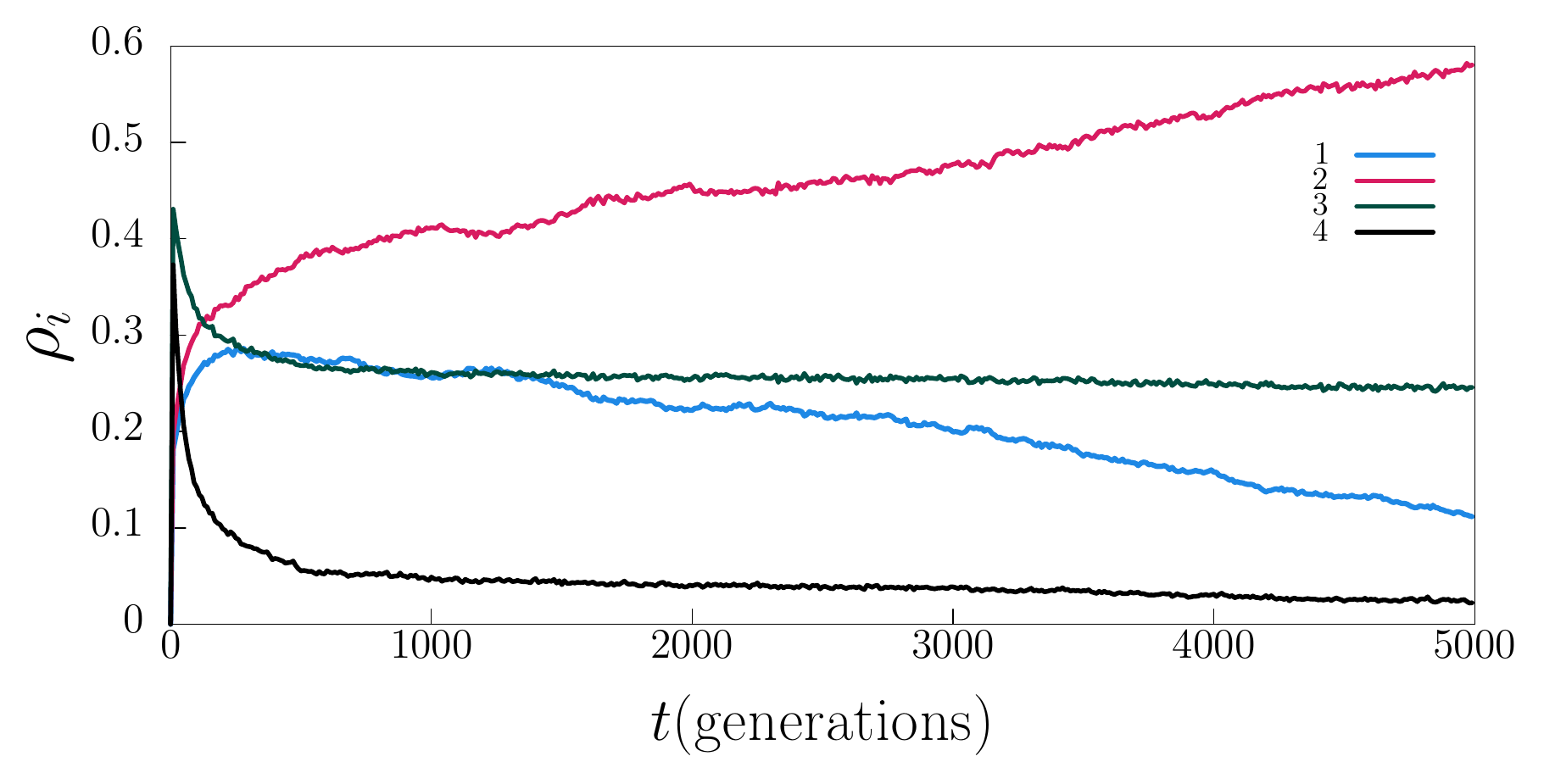}
\caption{Temporal changes in the spatial species densities during the simulation showed in Fig. \ref{fig2}. Blue, red, and green lines represent densities of prey species $1$, $2$, and predator spatial densities. The black line shows the dynamics of the densities of the predator with maximum predation capacity.}
	\label{fig3}
\end{figure}
\section{Results} \label{Sec3}
We first ran a single simulation in a lattice with $500^2$ grid points during a timespan of $5000$ generations, using the set of parameters: $m=0.1$, $r=0.9$, $p=0.78$, $d=0.12$, $\nu=0.5$, and $R=4$. Figure \ref{fig2} shows snapshots of the spatial patterns captured from the simulation: Figs. \ref{fig2a}, \ref{fig2b}, \ref{fig2c}, and \ref{fig2d} show the initial conditions and spatial configurations after $200$, $500$, $1000$, and $4000$ generations. The colours follow the scheme in Fig.~\ref{fig1}, where blue and red dots represent spatial positions occupied by only one individual of prey species $1$ or prey species $2$, respectively; yellow dots show the grid sites shared by a pair of individuals of each type of prey. Green dots represent the predators, whereas a white dot shows an empty space.  
The dynamics of the spatial configuration during the entire simulation is shown in https://youtu.be/9tDxXTkdePM;  the temporal change in the species densities are depicted in Fig \ref{fig3}.

At the very beginning of the simulation, the distribution of predators and individuals of both types of prey species is globally homogeneous (Fig.~ \ref{fig2a}). This promotes a high predation rate everywhere, resulting in fast growth in the predator's population, as shown in Fig.~\ref{fig3}. 
After that, departed regions occupied by organisms of one type of prey species arise, as depicted in Fig.~\ref{fig2b} by blue (prey species 1) and red (prey species 2) dots - we refer to the areas as single-prey domains. Due to the topological features of the two-dimensional space, organisms of both types of prey are present on the boundaries of the single-prey domains.
Because predators on the borders of single-prey domains 
consume organisms of both prey species, their chance of preying is higher than other predators within single-prey areas; thus, their predation capacity is maximum.
This is responsible for the interfaces with high predator density on the boundaries of the single-prey species, as shown by the orange dots in Fig.~\ref{fig2b}. Moreover, whenever a predator in the interface dies, an empty space is created, allowing prey to reproduce. In summary, a cyclic process ensures the stability of the interfaces: i) predators consume both types of prey and reproduce; ii) predators die, creating empty spaces; iii) new individuals of both types of prey fill the empty spaces. As time passes, the coarsening of the interface network causes the collapse of single-prey domains, as depicted in Fig.~\ref{fig2c} and Fig.~\ref{fig2d}. The dominance of prey species $2$ in Fig.~\ref{fig2e} results from the specific initial conditions in Fig.~\ref{fig2a}. Running realisations starting from different random initial conditions, we found that both prey species are equally likely to survive at the end of the simulations because the interaction probabilities are the same.

\section{Scaling regime} \label{Sec4}
 \begin{figure}
\centering
    \begin{subfigure}{.2\textwidth}
        \centering
        \includegraphics[width=35mm]{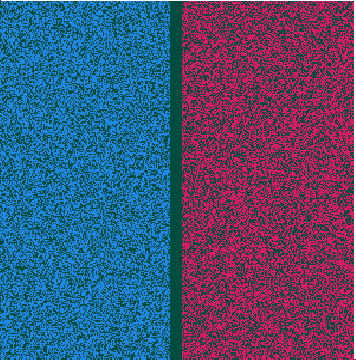}
        \caption{}\label{fig4a}
    \end{subfigure}
       \begin{subfigure}{.2\textwidth}
        \centering
        \includegraphics[width=35mm]{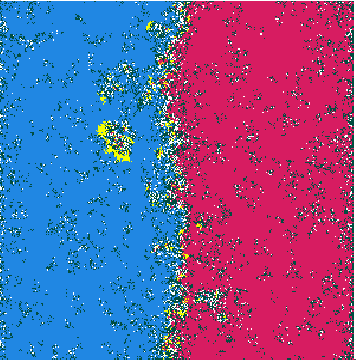}
        \caption{}\label{fig4b}
    \end{subfigure}\\
   \begin{subfigure}{.2\textwidth}
        \centering
        \includegraphics[width=35mm]{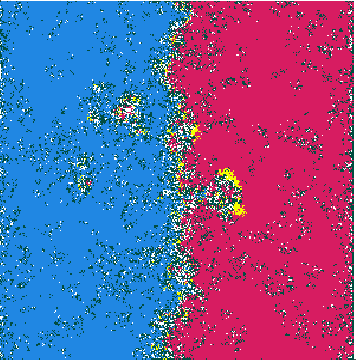}
        \caption{}\label{fig4c}
    \end{subfigure} 
       \begin{subfigure}{.2\textwidth}
        \centering
        \includegraphics[width=35mm]{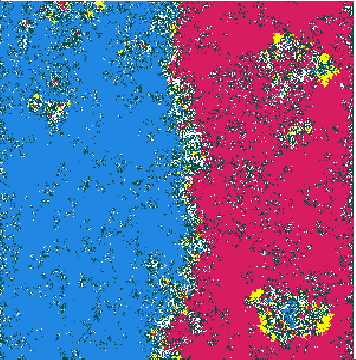}
        \caption{}\label{fig4d}
    \end{subfigure}
\caption{Snapshots captured from a simulation in a $300^2$ lattice with fixed boundary conditions, running until $400$ generations. The colours follow the same scheme of Fig.~\ref{fig1}.
Figures ~\ref{fig4a}, ~\ref{fig4b}, ~\ref{fig4c}, ~\ref{fig4d} depict the initial conditions and configurations after $76$, $118$, and $173$ generations. The video https://youtu.be/bdL2150lKfY shows the dynamics of the spatial configuration during the entire simulation.}
  \label{fig4}
\end{figure}

In the previous section, we found that single-prey domains surrounded by interfaces composed mainly of predators are formed from random initial conditions.
Figures~\ref{fig2f}, \ref{fig2g}, ~\ref{fig2h}, ~\ref{fig2i}, and ~\ref{fig2j} highlight the interface network in 
Figs.~~\ref{fig2a}, ~\ref{fig2b}, ~\ref{fig2c}, ~\ref{fig2d}, and ~\ref{fig2e}, respectively. Video https://youtu.be/bdL2150lKfY depicts the dynamics of the interface network.
The black dots depict predators
consuming organisms of both types of prey; empty spaces, single-prey domains, two-prey areas, and grid sites with predators feeding only on one type of prey were left uncoloured (white regions). 
The emerging interface network is similar to those studied previously in scenarios with interference competition of two species, where interfaces are composed of empty sites \cite{Avelino-PRE-86-031119,Pereira}. In our model, however, species segregation is not caused by direct competition but by the existence of a common predator. This means that the extinction of prey species $2$ occurs due to a coarsening process: some territories occupied by a single prey species grow, causing the collapse of other single-prey domains.  

\begin{figure*}
\centering
 \begin{subfigure}{.19\textwidth}
        \centering
        \includegraphics[width=34mm]{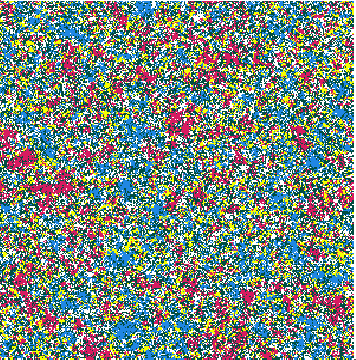}
        \caption{}\label{fig5a}
    \end{subfigure} %
    \begin{subfigure}{.19\textwidth}
        \centering
        \includegraphics[width=34mm]{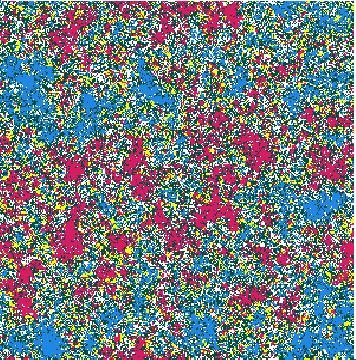}
        \caption{}\label{fig5b}
    \end{subfigure} %
       \begin{subfigure}{.19\textwidth}
        \centering
        \includegraphics[width=34mm]{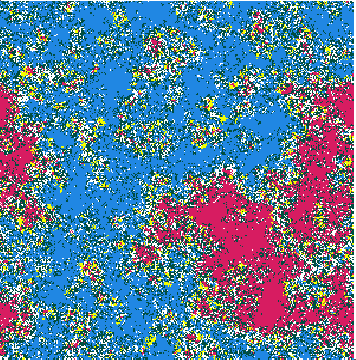}
        \caption{}\label{fig5c}
    \end{subfigure} %
   \begin{subfigure}{.19\textwidth}
        \centering
        \includegraphics[width=34mm]{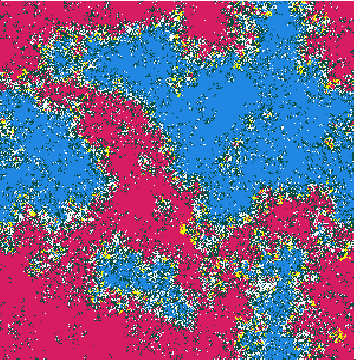}
        \caption{}\label{fig5d}
    \end{subfigure} %
    \begin{subfigure}{.19\textwidth}
        \centering
        \includegraphics[width=34mm]{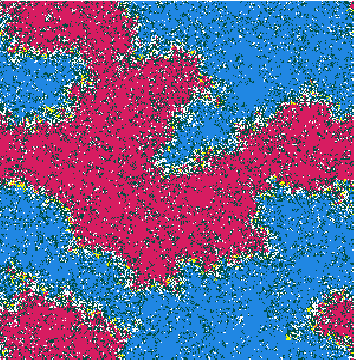}
        \caption{}\label{fig5e}
    \end{subfigure}
        \caption{Spatial patterns for various predator mortality probabilities. Figures ~\ref{fig5a}, ~\ref{fig5b}, ~\ref{fig5c}, ~\ref{fig5d} and ~\ref{fig5e} show snapshots taken from simulations in a grid with $300^2$ sites, after $500$ generations, for $d=0.17$, $d=0.16$, $d=0.15$, $d=0.14$,  and $d=0.13$, respectively.}
        \label{fig5}
\end{figure*}
To understand the dynamics of the spatial patterns, we study how the interface network changes in time. First, suppose an interface with curvature radius $r_{\kappa}$ and thickness $\epsilon\,\ll\,r_{\kappa}$, separating two distinct single-prey domains - for example, prey species $1$ outside de interface and prey species $2$ in the inner region. As predation interactions occur if the prey is present in the immediate predator's neighbourhood, the number of prey a predator can catch from outside the interface is proportional to $r_{\kappa} + \epsilon/2$ (the outer interface length). In contrast, the number of prey from the single-prey domain inside the border is proportional to $r_{\kappa} - \epsilon/2$ (the inner interface length). 
Because of this, the difference between the average number of prey consumed per unit time from outside and inside the interface is proportional to $\epsilon$: the time necessary to consume the organisms of prey $2$ (inside the interface) is smaller than all individuals organisms of prey $1$ (outside the interval). Thus, as time passes, predation of prey outside and inside the border provokes the reduction of the interface curvature radius.

The time necessary for the circular single-prey domain radius to be reduced by $\Delta r_{\kappa} \ll r_{\kappa}$ is proportional 
to the interface length, i.e., proportional to $r_{\kappa}$. 
This means that the dynamics of the spatial pattern network is curvature-driven, leading to
collapses of single prey spatial domains with a velocity proportional to its curvature, which is typical of non-relativistic interfaces in condensed matter \cite{Joana1,Joana2}.

Now, let us define the network characteristic length at time $t$ as
\begin{equation}
L(t)\,=\,\sqrt{\frac{A}{N_D(t)}}\,=\,\sqrt{\frac{\mathcal{N}}{N_D(t)}} \label{eeq1}
\end{equation} 
where $A$ is the total grid area, that is equal to the total number of grid points $\mathcal{N}$; $N_{D} (t)$ is the number of single-prey domains in the time $t$. 
As demonstrated in Ref.~\cite{Avelino-PRE-86-031119} for curvature driven interface networks without junctions, the number of single-prey spatial domains, $N_D$, decreases in time according to 
\begin{equation}
N_D(t)\,=\, \frac{C}{t} \label{eqq1}
\end{equation}
where $C$ is a positive constant. As a consequence, 
the characteristic length
increases with time according to scaling law
\begin{equation}
L\,\propto t^{1/2}. \label{eq1}
\end{equation}
Eq.~\ref{eq1} describes the dynamics of the interface networks in our apparent competition model, as observed in Figs.~\ref{fig2f} to \ref{fig2j}.

Now, we aim to write Eq.~\ref{eq1} is terms of the total number of predators forming interfaces between single-prey domains. We define the total number of predators with maximum predation capacity as $I_P$ (predators forming interfaces between single-prey domains); thus,  $I_P\,=\,\epsilon\,L_T$
where $L_T$ is total interface length and $\epsilon$ the interface thickness. Given that the interface thickness is defined by the mobility probability \cite{Roman}, once the model parameters are defined, the interface thickness is constant in time and space, leading to
\begin{equation}
I_P\,\propto\, L_t. \label{eqt}
\end{equation}

The number of single-prey domains is given by
\begin{equation}
N_D\,=\,\frac{L_T}{P_D}, \label{eqr}
\end{equation}
where $P_D$ is the average domain perimeter. Given that the network characteristic length $L$ is defined by the average domain radius, one has 
\begin{equation}
P_D\,\propto\, L, \label{eqe}
\end{equation}
which allows Eq.~\ref{eqr} to be written as 
\begin{equation}
N_D\,\propto\,\frac{L_T}{L}. \label{eqr}
\end{equation}  

Finally, using $N_D\,\propto\,L^{-2}$ (Eq.~\ref{eeq1}), we combine 
Eqs.~\ref{eqt} to \ref{eqr}, to find that
\begin{equation}
I_P\,\propto\,L^{-1}. \label{eqi}
\end{equation}
Substituting Eq.~\ref{eq1} in Eq.~\ref{eqi}, we write the scaling power law that that characterises the dynamics of the interface networks in our apparent competition model as
\begin{equation}
I_{P}\,\propto\,t^{-1/2}. \label{eqo}
\end{equation}

To verify the analytical prediction to the scaling law given by Eq.~\ref{eqo}, we performed a set of $100$ simulations in lattices with $300^2$ sites, with a timespan of $1000$ generations. Each simulation started from different random initial conditions and ran for the same set of parameters of the realisation shown Fig.~\ref{fig2}. We calculated how $I_P$ decreases with time considering the function $I_P \propto t^{-\lambda}$, where $\lambda$ is the scaling exponent. Our outcomes show that $\lambda = 0.492338 \pm 0.0002813$, which is very close to the theoretical prediction in Eq.~\ref{eqo}, confirming that the dynamics of the interface network attains the expected scaling regime.
\section{The role of predator mortality} \label{Sec5}
To understand the role of predator mortality on the pattern formation and coarsening dynamics of the spatial patterns in our apparent competition model, we first run a single simulation with the prepared initial conditions depicted in Fig.~\ref{fig4a}. 
Different prey species are put on opposite grid sides in this initial configuration. In contrast, predators are randomly distributed within the single-prey domain or placed in the central vertical interface. As we are interested in observing the dynamics of the central vertical interface, we relaxed the periodic boundary conditions in this cases; thus, we assumed a lattice with fixed edges. The parameters are the same as in the previous simulations except for $d\,=\,0.14$ - consequently, $p\,=\,0.76$.

The outcomes reveal that although the passage of prey to the opposite domain is rare, a prey finds plenty of room to reproduce when it occurs. For example, whenever an individual of prey species $2$ reaches the territory with prey species $1$, it reproduces because the prey species do not compete for space. This process leads to the formation of waves of two-prey regions, predators with maximum predation capacity, and empty spaces resulting from predator death, as depicted in Fig.~\ref{fig4b}.
The waves propagate into the single-prey domains until the stochasticity of predation-prey interactions leads to the consumption of all individuals of one type of prey. In this situation, predators can no longer prey at the maximum rate; consequently, part of the predators die, leaving a spatial concentration of empty space filled by the remaining prey. According to Figs.~\ref{fig4c} and \ref{fig4d}, two scenarios are possible: i) if the remaining prey belongs to the same prey species which dominates the spatial domain where the waves were travelling, the waves disappear without a trace; ii) otherwise, a new spatial domain emerges, surrounded by a closed interface; the new domain further collapses due to interface curvature.

Compared with the simulation shown in Fig.~\ref{fig2}, the predator density is lower because: 
\begin{enumerate}
\item \label{pp}
The higher predator mortality yields the appearance of a larger number of empty spaces per time unit, allowing a higher reproduction rate of both types of prey in the interface. 
\item \label{qq}
The lower predation probability results in fewer organisms of both prey species devoured, and consequently, fewer births of new predators per unit time, promoting the increase of prey densities in the interface.
\end{enumerate}
Both \ref{pp} and \ref{qq}  results in a higher prey density in the interface.
This process let a small number of organisms cross the interface, creating waves that can impact the coarsening process observed in the previous section. To quantify how the presence of the waves alters the dynamics of the spatial patterns demonstrated in Sec.~\ref{Sec4}, we performed simulations for various values of $d$ and $p$ starting from random initial conditions. We maintain the interface thickness unaltered in all simulations by assuming the same mobility probability $m=0.1$ for all organisms of every species; for predators, $d$ and $p$ vary obeying the constraining $d+p\,=\,0.9$ (for reasons of simplification, our numerical analysis will focus on the parameter $d$). All simulations were performed in lattices with $300^2$ sites, starting from the same random initial conditions. Snapshots showing the final spatial configuration are depicted in Fig.~\ref{fig5}.

\begin{figure}
\centering
        \includegraphics[width=75mm]{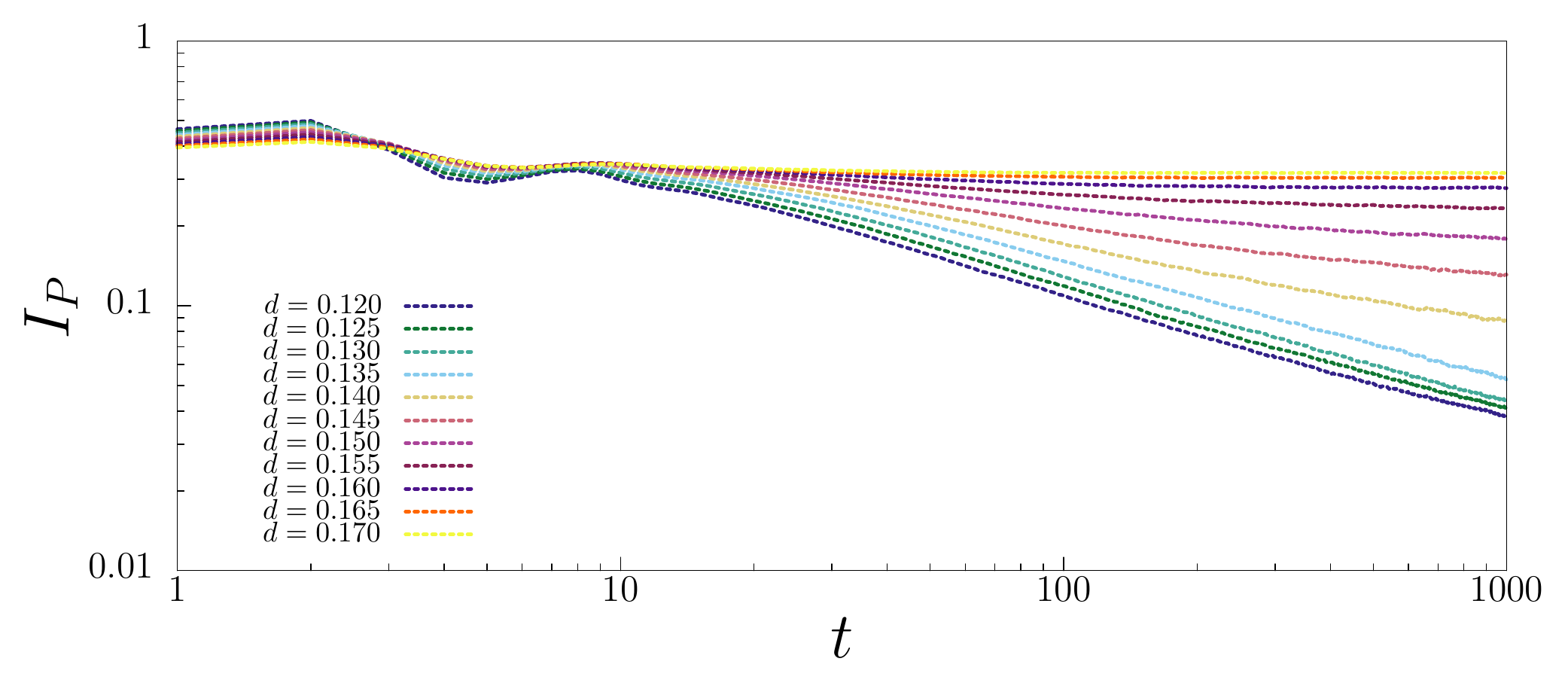}
\caption{Temporal variation of total number of predators with maximum predation capacity for various predator mortality probabilities.
The results were obtained from simulations running in lattices with $300^2$ sites.}
	\label{fig6a}
\end{figure}
\begin{figure}
\centering
        \includegraphics[width=75mm]{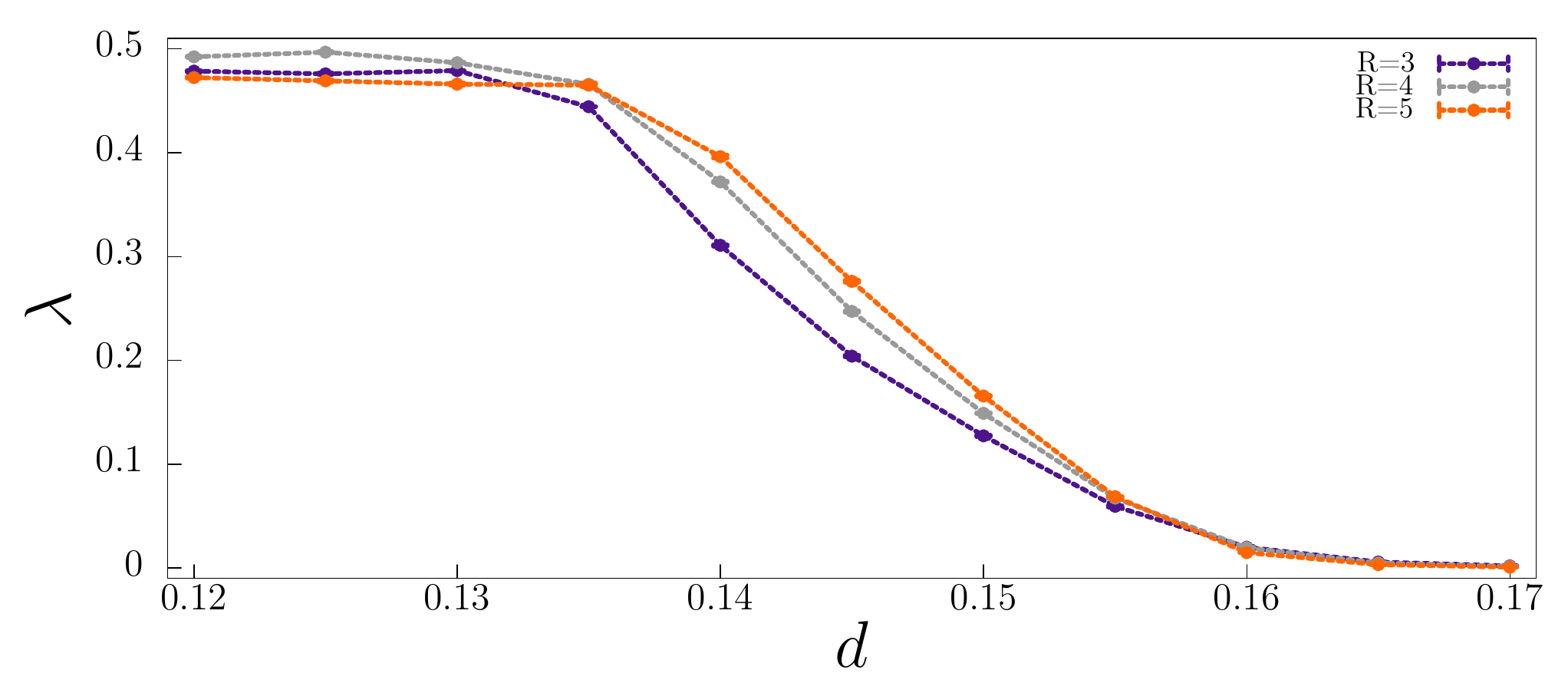}
\caption{Scaling exponent in terms of the predator mortality. The purple, grey, red and orange lines show how the dynamics of the spatial patterns of the changes with $d$, for the frequency radius $R=3$, $R=4$, and $R=5$, respectively.}
	\label{fig6b}
\end{figure}
\begin{table}[ht]
\caption{Scaling exponent crossover parameters} 
\centering 
\begin{tabular}{c c c} 
\hline\hline 
$R$ & $\alpha$ & $\gamma$ \\ [0.5ex] 
\hline 
$3$ & $95.032 \pm 6.103$ & $0.143536 \pm 0.0003827$  \\
$4$ & $104.09 \pm 4.255$ & $0.14541 \pm 0.0002227$  \\
$5$ & $103.385  \pm  6.746$ & $0.14623 \pm 0.0003578$  \\ [1ex] 
\hline 
\end{tabular}
\label{tab1} 
\end{table}

Let us first observe the spatial configuration depicted in Figs.~\ref{fig5a}, for $d=0.17$. In this case, a higher concentration of empty spaces in the grid and a lower predator density stimulate the arising of irregular two-prey domains everywhere (yellow areas). 
As $d$ decreases, regions inhabited by two types of prey are less common because of the presence of more predators throughout the lattice, as depicted in Figs.~\ref{fig5b} and ~\ref{fig5c}, for $d=0.16$ and $d=0.14$, respectively. 
Two-prey areas are quickly eliminated in the lower predator mortality simulations because of the high concentration of predators with maximum predation capacity. This results in departed single-prey domains depicted in Figs.~\ref{fig5d}, for $d=0.14$, where blue and red regions show the areas inhabited mainly by prey $1$ and $2$, respectively. However, there is a slight chance of a prey managing to reach the opposite domain, creating the waves observed in Fig.~\ref{fig4}.
The frequency of the arising of the waves is reduced for $d=0.13$, as depicted in Fig.~\ref{fig5e}, where the formation of two-prey regions (small yellow areas) roughens the interface without propagating on the single-prey domains. In this case, whenever a prey moves towards the interface, the probability of being killed is high, limiting individuals of different prey species to live in separate domains.

Therefore, the lower the predator mortality is, the less the dynamics of the spatial patterns are affected by the noise introduced by waves spreading through the single-prey species observed in Fig. ~\ref{fig4}. This is similar to the thermal fluctuations that decelerate the two-dimensional coarsening 
of in nonequilibrium spin models \cite{Topics, Arezon1,Interplay,Arezon2,Oliveira1,Tartaglia_2018}. It has been shown that the thermal fluctuations are responsible for the roughening of the interface (opposing to the curvature-driven growth), and the arising of thermal equilibrium domains that are not related to the coarsening mechanism.

Here, the fluctuations are controlled by predator mortality. Thus, to investigate how $d$ impacts the dynamics of the total number of predators with maximum 
predation capacity, we ran sets of $100$ simulations in lattices with $300^2$ sites, running until $1000$ generations, for several values of $d$ in the interval $0.12\,\leq\,d\,\leq\,0.17$. Figure~\ref{fig6a} shows the time dependence of $I_P$ for various predator mortality, for $R=3$. 
Observing the lines depicting the temporal change of $I_P$ for $d=0.12$ (dark blue), $d=0.125$ (dark green), $d=0.13$ (light green), $d=0.135$ (light blue), $d=0.14$ (gold), $d=0.145$ (light pink), $d=0.15$ (dark pink), $d=0.155$ (ruby), $d=0.16$ (purple), $d=0.165$ (orange), and $d=0.17$ (yellow), one concludes that
the dynamics of spatial patterns is strongly dependent on the predator mortality.
For high $d$, the dynamics is not curvature-driven; thus, the definition of the characteristic length discussed in the previous section is valid only for values near to $d=0.12$.

\begin{figure}
\centering
 \begin{subfigure}{.45\textwidth}
        \centering
        \includegraphics[width=75mm]{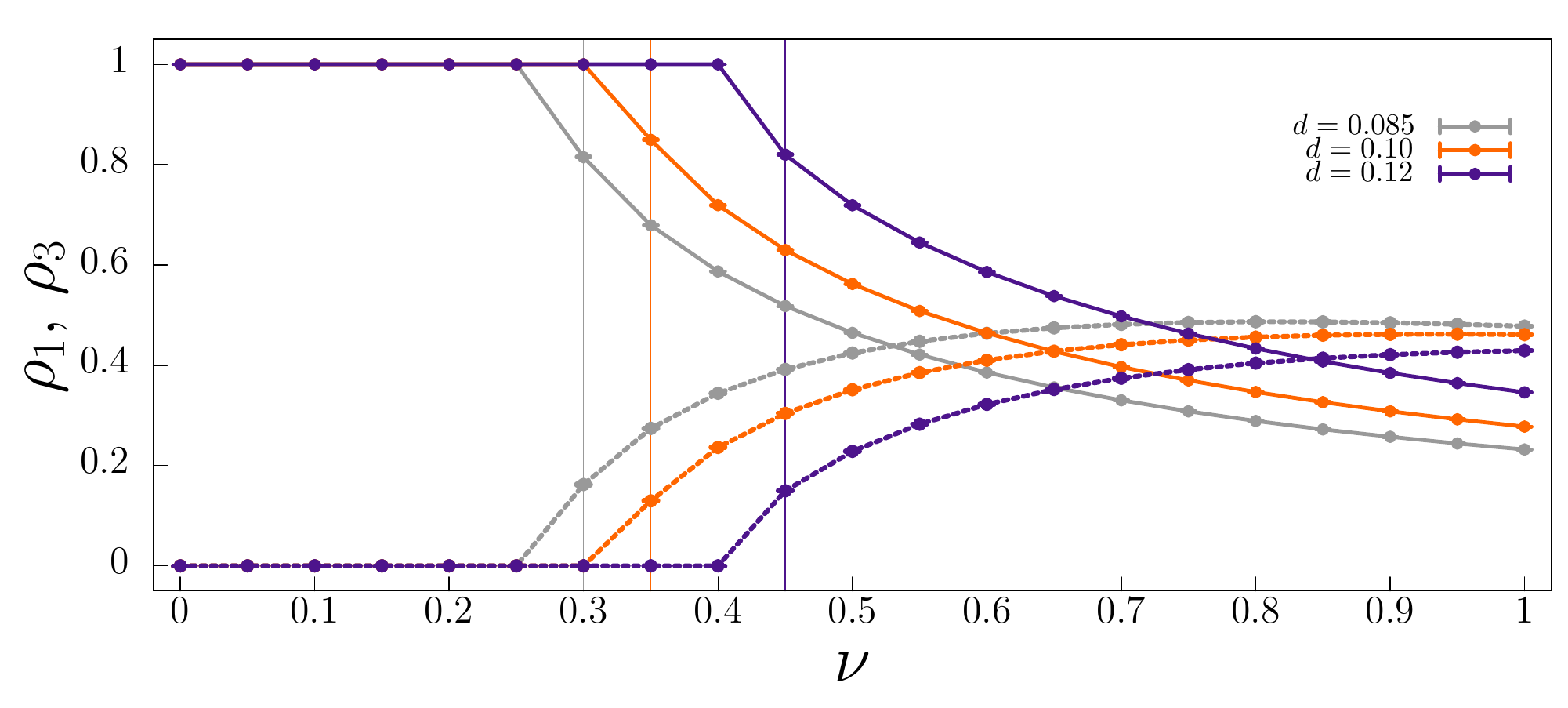}
        \caption{}\label{fig7a}
    \end{subfigure} \\%
    \centering
 \begin{subfigure}{.45\textwidth}
        \centering
        \includegraphics[width=75mm]{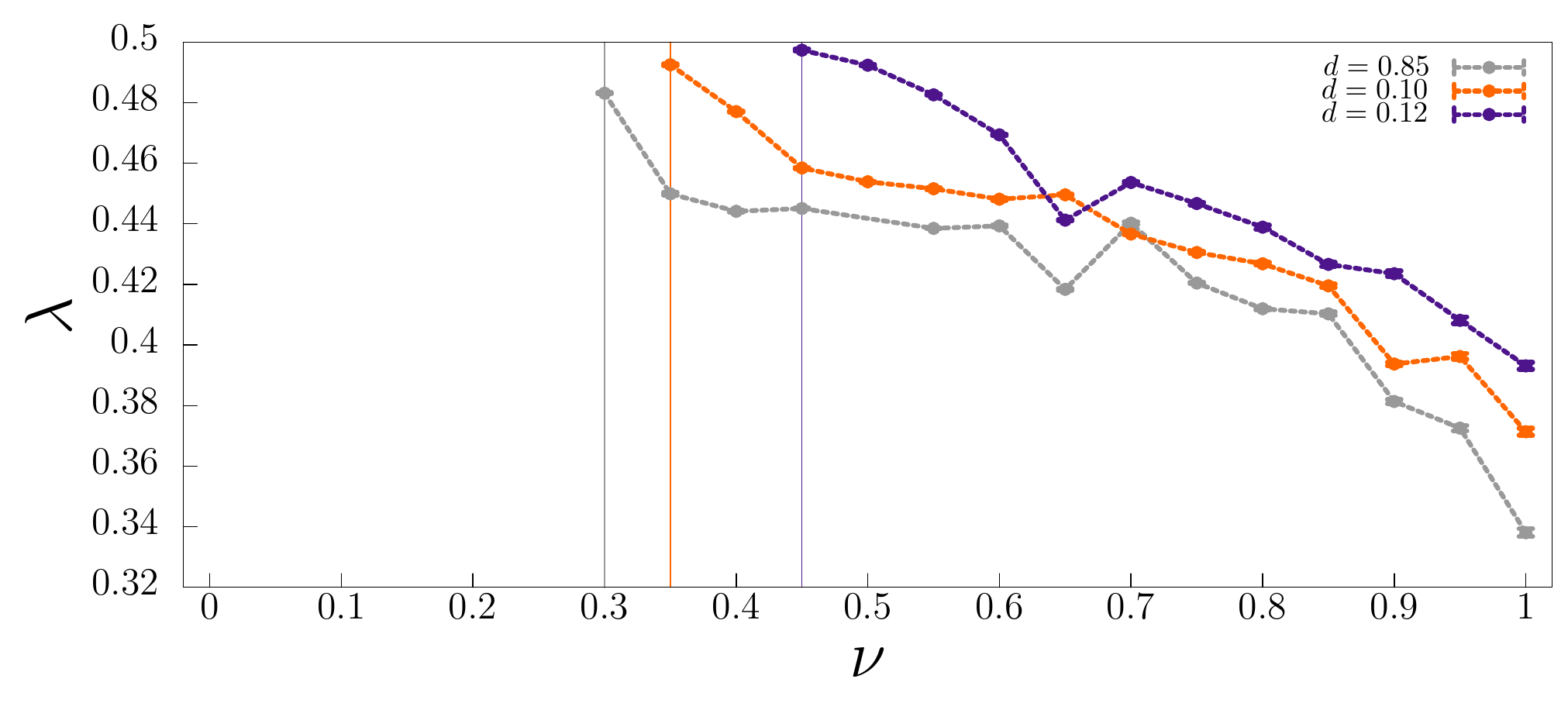}
        \caption{}\label{fig7b}
    \end{subfigure} %
\caption{Predator and prey densities in single-prey domains (Fig.~\ref{fig7a}) and scaling exponent (Fig.~\ref{fig7b}) in terms of the predation capacity in single-prey domains. The grey, orange, and blue lines depict the results for predator mortality $d\,=\,0.10$, $d\,=\,0.12$, and $d\,=\,0.15$, respectively. The error bars show the standard deviation of the results averaged from $100$ simulations. The vertical lines indicate the minimum predation capacity for predators to persist in single-prey areas.}
	\label{fig7}
\end{figure}

To quantify the changes in the coarsening dynamics, we studied how the total number of predators with maximum predation capacity changes with time for several values of $d$ in the interval $0.12\,\leq\,d\,\leq\,0.17$. For this purpose, we consider that 
\begin{equation}
I_P \propto t^{-\lambda(d)}, \label{eqcross}
\end{equation}
where $\lambda(d)$ is a function of the predator mortality.
We have not considered the first $50$ generations to calculate the scaling exponent to avoid the transient interface network formation fluctuations observed in Fig.\ref{fig6a}. Moreover,
as the definition of the predation capacity depends on the frequency radius $R$, we repeated the simulations for $R=3,4,5$. 

Figure~\ref{fig6b} shows the average value of the scaling exponent $\lambda$ computed from $100$ realisations running until $1000$ generations in lattices with $300^2$ grid points. The purple, grey, and orange lines depict $\lambda$ as function of $d$ for $R=3$, $R=4$, and $R=5$, respectively. The error bars show the standard deviations.
The outcomes show that for $d \approx 0.12$, the coarsening dynamics attains the regime scaling predicted in Eq. \ref{eqo} since the impact of the fluctuations is reduced for low predator mortality. On the contrary, for $d \approx 0.17$, the average total number of predators with maximum predation capacity remains approximately constant with time. This happens because growing
single-prey species domains are not formed, promoting coexistence among predator, prey $1$ and prey $2$, as we observed in Fig.~\ref{fig5a}.

Our outcomes show a crossover between two limit scenarios: curvature driven network coarsening ($\lambda \approx 0.5$) and the coexistence regime ($\lambda \approx 0.0$). We then described the crossover employing the function
\begin{equation}
\lambda = 0.25\,(1-\,\tanh \left[\alpha\,(d\,-\,\gamma)\right], \nonumber
\end{equation}
to find the best fit of the results presented in Fig.~\ref{fig6b}. The parameters $\alpha$ and $\gamma$, which are functions of $R$, appear in Table~\ref{tab1}. Our numerical results also demonstrate that the coarsening of the interface network does not depend significantly on the frequency radius $R$.

\section{The influence of the predation capacity}\label{Sec6}

The spatial predator-prey interactions determine the population dynamics within the single-species spatial domains. Now we explore how predation capacity controls predator and prey local densities in regions far from the interfaces. 
To perform this investigation, we run two groups of simulations: 
\begin{itemize}
\item
We simulated a scenario where only organisms of one type of prey and predators are present in the grid - since both types of prey provide the same predation capacity, we chose prey $1$ in this numerical experiment. Initially, individuals were distributed randomly in the lattice, each species occupying one-third of the grid (one-third of the grid was left empty). The goal was to find the minimum predation capacity, $\nu^{\ast}$, that allows to predator to survive in patches far from the boundaries of the single-prey domains. Also, we aim to calculate the predator and prey populations in terms of $\nu$.
\item
To generalise the results presented in the previous sections, where predation capacity within single-prey spatial domains was reduced to half the value inside two-prey areas, we run the one-predator two-prey simulations for $\nu^{\ast}\,\leq\,\nu\, \leq\,1$. We quantified the effects of the local predator-prey dynamics within single-prey domains by calculating the dependence of the scaling power law on $\nu$, for various $d$.
\end{itemize}

We ran groups of $100$ simulations in grids with $300^2$ sites, with a timespan of $1000$ generations for $0\,\leq\,\nu\, \leq\,1$. We repeated the simulations for $d=0.085$, $d=0.10$, and $d=0.12$, with $m=0.1$ and $R=4$. The results are shown in Fig.~\ref{fig7}, where the purple, grey and orange lines depict the results for $d\,=\,0.10$, $d\,=\,0.12$, and $d\,=\,0.15$, respectively.
According to the outcomes in Fig.~\ref{fig7a}, the minimum predation capacity $\nu^{\ast}$ necessary for a predator to survive in single-prey domains
depends on $d$. The dashed and solid lines show how prey and predator densities, $\rho_1$ and $\rho_3$, change in terms of $\nu$: as predation capacity grows, local predator density nonlinearly increases in single-prey spatial domains. 

We then investigated the role of the predation capacity in our apparent competition model by calculating a generalised scaling exponent in the interval $\nu^{\ast}\,\leq\,\nu\,\leq\,1$, for $d=0.0875$, $d=0.10$, and $d=0.12$.
The results depicted in Fig.~\ref{fig7b} reveal that the scaling exponent reaches the maximum value for the minimum predation capacity that allows the persistence of the predator in case of only one type of prey exists. Overall, the findings indicate that for $\nu^{\ast}$, the interface network approaches the scaling regime theoretically predicted in Sec.~\ref{Sec3}. 
As $\nu^{\ast}$ represents the scenario where predators have less chance to reproduce within single-prey areas, waves are less likely to propagate throughout single-prey domains. Therefore, for $\nu^{\ast}$ the effects of the noise in the interface network is minimum, resulting in a
scaling power law that approaches the theoretical prediction $\lambda = 1/2$.

\section{Conclusions} \label{Sec7}
We study the formation and dynamics of spatial patterns in apparent competition spatial models. We focus on the case of two species living without any interference competition but sharing a common predator.
We performed a series of stochastic numerical simulations considering that predation capacity is higher in spatial regions where both types of prey are consumed. 

Our results show that areas inhabited mainly by only one kind of prey arise, with predators concentrated on the borders of these single-prey domains. The interfaces separating single-prey domains are mainly formed by predators consuming both types of prey. 
We demonstrated that the dynamics of the interface network is curvature-driven, predicting that the scaling regime for the coarsening dynamics leads to power-law function: $I_P \propto t^{-1/2}$, where
$I_P$ is the total number of predators consuming individuals of both types of prey. Studying a case where organisms of different types of prey are limited to living in departed patches, we confirmed that the interface network attains a scaling regime with the exponent scaling very close to the theoretical prediction. This outcome shows that the spatial pattern network in our apparent competition model changes in time according to the same standard scaling law found in other areas of nonlinear science.

Our findings also revealed that pattern formation is strongly affected by predator density. Fixing the mobility probability $m$ in our stochastic simulations, we 
explored the effects of the predator density by varying the predator mortality $d$ and the predation probability, keeping the constraining $p+d=1-m$.
We found that the higher the predator mortality (the lower the predation probability) is, the less probable the formation of growing departed single-prey domains. In this case, the system to depart the theoretical scaling regime $I_P \propto t^{-1/2}$ to a coexistence regime defined by $I_P \propto t^{0}$. Running many simulations for a wide range of predator mortality, we analytically described the crossover between the asymptotic regimes.

Finally, we investigated the role of the predation capacity in the local population dynamics within the single-prey dynamics. Our findings show that a minimum predation capacity is necessary to guarantee survival for given predator mortality. We verified that the persistence of predators in regions far from the borders of the single-prey domains is crucial to the scaling regime attained by the interface network: ii) if predation capacity is the minimum necessary to ensure predator persistence, distinct prey species are limited to live within the single-prey domains; this prevents the formation of waves, resulting in a scaling regime whose exponent agrees with the theoretical prediction; ii) if predation capacity is higher than the minimum necessary to ensure coexistence within single prey domains, predator population grows; the lower prey concentration in the single-prey domains brings reduction on the predator-prey activity, and a consequent reduction in the exponent scaling that describe the coarsening dynamics.

Foraging species are observed in many biological systems, where directional movement is motivated by the search for patches with high prey density \cite{Motivation1,Motivation2}.
Our investigation can be generalised to include a prey-taxis movement, where a predator can scan their environment to choose the best direction to move, instead of walking randomly on the grid \cite{rev31,rev32}. In this case, predators efficiently arrive at areas with both types of prey; thus, the effect of the apparent competition on the pattern formation is reinforced. Moreover, once the single-prey regions are formed, prey-taxis movement increases predator concentration in the interfaces, where predation capacity is maximum. This may reduce the chances of prey crossing the interface to the opposite domain; thus, reducing the fluctuations that decelerate the coarsening dynamics.

To the best of our knowledge, this is the first time the dynamics of interface networks is quantified in apparent competition models.
Our results may be generalised to investigate other topological patterns well known in systems with direct competition, for example, interface with junctions or string networks \cite{Avelino-PRE-86-036112,Avelino-PRE-86-031119,Pereira,PhysRevE.89.042710,PhysRevE.99.052310,Roman,strings1,strings2}. 
The main difference is that, in our model, the topological interface networks are formed mostly by predators instead of empty spaces. 
For example, in systems where three or four prey species share a common predator, departed single-prey domains appear. If predation capacity is the same in single-prey territories, irrespective of the prey species, the interfaces meet in $Y$-type junctions (\cite{Avelino-PRE-86-036112}). The results may be helpful for biologists to understand the mechanisms that control the dynamics of spatial patterns in systems with the apparent competition.

\vspace{0.5cm}
We thank Arne Janssen for enlightening discussions. We acknowledge ECT, Fapern/CNPq, IBED, and IIN-ELS for financial and technical support.

\bibliography{ref}

\begin{thebibliography}{10}
\expandafter\ifx\csname url\endcsname\relax
  \def\url#1{\texttt{#1}}\fi
\expandafter\ifx\csname urlprefix\endcsname\relax\def\urlprefix{URL }\fi
\expandafter\ifx\csname href\endcsname\relax
  \def\href#1#2{#2} \def\path#1{#1}\fi

\bibitem{ecology}
M.~Begon, C.~R. Townsend, J.~L. Harper, Ecology: from individuals to
  ecosystems, Blackwell Publishing, Oxford, 2006.

\bibitem{nowak06evolutionaryDynamicsBOOK}
M.~A. Nowak, Evolutionary Dynamics: Exploring the Equations of Life, Harvard
  University Press, 2006.

\bibitem{Nature-bio}
A.~Purvis, A.~Hector, Getting the measure of biodiversity, Nature 405 (2000)
  212--2019.

\bibitem{Coli}
B.~Kerr, M.~A. Riley, M.~W. Feldman, B.~J.~M. Bohannan, Local dispersal
  promotes biodiversity in a real-life game of rock–paper–scissors, Nature
  418 (2002) 171.

\bibitem{bacteria}
B.~C. Kirkup, M.~A. Riley, Antibiotic-mediated antagonism leads to a bacterial
  game of rock-paper-scissors in vivo, Nature 428 (2004) 412--414.

\bibitem{Allelopathy}
R.~Durret, S.~Levin, Allelopathy in spatially distributed populations, J.
  Theor. Biol. 185 (1997) 165--171.

\bibitem{lizards}
B.~Sinervo, C.~M. Lively, The rock-scissors-paper game and the evolution of
  alternative male strategies, Nature 380 (1996) 240--243.

\bibitem{Extra1}
I.~Volkov, J.~R. Banavar, S.~P. Hubbell, A.~Maritan, Patterns of relative
  species abundance in rainforests and coral reefs, Nature 450 (2007) 45.

\bibitem{But1}
G.~Lei, I.~Hanski, Spatial dynamics of two competing specialist parasitoids in
  a host metapopulation, Journal of Animal Ecology 67~(3) (1998) pp. 422--433.

\bibitem{But2}
M.~Zalucki, R.~Kitching, Oecologia 53~(2) (1982) 201--207.

\bibitem{Chesson}
P.~Chesson, Updates on mechanisms of maintenance of species diversity, Journ.
  of Ecology 106 (2018) 1773--1794.

\bibitem{Bonsall}
M.~B. Bonsall, M.~P. Hassell, Apparent competition structures ecological
  assemblages, Nature 388 (1997) 371--373.

\bibitem{Taylor}
J.~M. Taylor, W.~E. Snyder, Are specialists really safer than generalists for
  classical biocontrol, BioControl 66 (2021) 9--22.

\bibitem{Holt}
R.~D. Holt, Predation, apparent competition, and the structure of prey
  communities, Theor. Popul. Biol. 12 (1977) 197--229.

\bibitem{Australis}
G.~P. Bhattarai, L.~A. Meyerson, J.~T. Cronin, Geographic variation in apparent
  competition between native and invasive phragmites australis, Ecology 98
  (2017) 3149--358.

\bibitem{Boreal}
B.~T. Neufeld, C.~Superbie, R.~J. Greuel, T.~Perry, P.~A. Tomchuk, D.~Fortin,
  P.~D. McLoughlin, Disturbance-mediated apparent competition decouples in a
  northern boreal caribou range, Wildlife Management 85 (2020) 254--270.

\bibitem{Field}
D.~M. Tompkins, R.~A.~H. Draycott, P.~J. Hudson, Field evidence for apparent
  competition mediated via the shared parasites of two gamebird species, Ecol.
  Letters 1 (2001) 10--14.

\bibitem{Hyperpredation}
E.~Caudera, V.~Simona, S.~Bertolino, J.~Cerri, E.~Venturino, A mathematical
  model supporting a hyperpredation effect in the apparent competition between
  invasive eastern cottontail and native european hare, Bull. of Math. Biol. 83
  (2021) 51.

\bibitem{ApparentEquations}
L.~D. Fernandes, M.~A.~M. Aguiar, Turing patterns and apparent competition in
  predator-prey food webs on networks, Phys. Rev. E 86 (2012) 056203.

\bibitem{MathematicalImpulsive}
H.~Yu, S.~Zhong, R.~P. Agarwalc, Mathematics and dynamic analysis of an
  apparent competition community model with impulsive effect, Math. and Comp.
  Model. 52 (2010) 25--36.

\bibitem{Reichenbach-N-448-1046}
T.~Reichenbach, M.~Mobilia, E.~Frey, Mobility promotes and jeopardizes
  biodiversity in rock-paper-scissors games, Nature 448 (2007) 1046--1049.

\bibitem{Avelino-PRE-86-036112}
P.~P. Avelino, D.~Bazeia, L.~Losano, J.~Menezes, B.~F. Oliveira, Junctions and
  spiral patterns in generalized rock-paper-scissors models, Phys. Rev. E 86
  (2012) 036112.

\bibitem{uneven}
J.~Menezes, B.~Moura, T.~A. Pereira, Uneven rock-paper-scissors models:
  Patterns and coexistence, Europhysics Letters 126~(1) (2019) 18003.

\bibitem{Moura}
B.~Moura, J.~Menezes, Behavioural movement strategies in cyclic models,
  Scientific Reports 11 (2021) 6413.

\bibitem{Anti1}
J.~Menezes, Antipredator behavior in the rock-paper-scissors model, Phys. Rev.
  E 103 (2021) 052216.

\bibitem{anti2}
J.~Menezes, B.~Moura, Mobility-limiting antipredator response in the
  rock-paper-scissors model, Phys. Rev. E 104 (2021) 054201.

\bibitem{neigh}
D.~Bazeia, M.~Bongestab, B.~{de Oliveira}, Influence of the neighborhood on
  cyclic models of biodiversity, Physica A: Statistical Mechanics and its
  Applications 587 (2022) 126547.

\bibitem{Avelino-PRE-86-031119}
P.~P. Avelino, D.~Bazeia, L.~Losano, J.~Menezes, von neummann's and related
  scaling laws in rock-paper-scissors-type games, Phys. Rev. E 86 (2012)
  031119.

\bibitem{Pereira}
T.~A. Pereira, J.~Menezes, L.~Losano, Interface networks in models of competing
  species, Intern. J. of Mod., Sim. and Sci. Comp. 9 (2018) 1850046.

\bibitem{PhysRevE.89.042710}
P.~P. Avelino, D.~Bazeia, L.~Losano, J.~Menezes, B.~F. de~Oliveira, Interfaces
  with internal structures in generalized rock-paper-scissors models, Phys.
  Rev. E 89 (2014) 042710.

\bibitem{PhysRevE.99.052310}
P.~P. Avelino, J.~Menezes, B.~F. de~Oliveira, T.~A. Pereira, Expanding spatial
  domains and transient scaling regimes in populations with local cyclic
  competition, Phys. Rev. E 99 (2019) 052310.

\bibitem{Joana1}
P.~Avelino, J.~Oliveira, C.~Martins, Understanding domain wall network
  evolution, Physics Letters B 610~(1) (2005) 1--8.

\bibitem{Joana2}
P.~P. Avelino, R.~Menezes, J.~C. R.~E. Oliveira, Unified paradigm for interface
  dynamics, Phys. Rev. E 83 (2011) 011602.

\bibitem{Roman}
A.~Roman, D.~Konrad, M.~Pleimling, Cyclic competition of four species: domains
  and interfaces, J. Stat. Mech. 7 (2012) P07014.

\bibitem{Topics}
L.~F. Cugliandolo, Topics in coarsening phenomenai, Physica A 389 (2010)
  4360–4373.

\bibitem{Arezon1}
A.~Sicilia, J.~J. Arenzon, A.~J. Bray, L.~F. Cugliandolo, Domain growth
  morphology in curvature-driven two-dimensional coarsening, Phys. Rev. E 76
  (2007) 061116.

\bibitem{Interplay}
P.~Roy, P.~Sen, Interplay of interfacial noise and curvature-driven dynamics in
  two dimensions, Phys. Rev. E 95 (2017) 020101.

\bibitem{Arezon2}
M.~P.~O. Loureiro, J.~J. Arenzon, L.~F. Cugliandolo, A.~Sicilia,
  Curvature-driven coarsening in the two-dimensional potts model, Phys. Rev. E
  81 (2010) 021129.

\bibitem{Oliveira1}
M.~J. F. F. S. M.~A. Oliveira, M.~J., Nonequilibrium spin models with ising
  universal behaviour, J. Phys. A: Math. Gen. 26 (1993) 2317.

\bibitem{Tartaglia_2018}
A.~Tartaglia, L.~F. Cugliandolo, M.~Picco, Coarsening and percolation in the
  kinetic 2d ising model with spin exchange updates and the voter model
  2018~(8) (2018) 083202.

\bibitem{Motivation1}
G.~S. Fraenkel, D.~L. Gunn, The orientation of animals. kineses, taxes, and
  compass reactions, The American Naturalist 75~(761) (1941) 604.

\bibitem{Motivation2}
D.~Kuefler, T.~Avgar, J.~M. Fryxell, Rotifer population spread in relation to
  food, density and predation risk in an experimental system, Journ. Anim.
  Ecol. 81~(2) (2012) 323.

\bibitem{rev31}
K.~Wang, Q.~Wang, F.~Yu, Stationary and time-periodic patterns of two-predator
  and one-prey systems with prey-taxis, Discrete \& Continuous Dynamical
  Systems 37~(1) (2017) 505--543.

\bibitem{rev32}
Q.~Wang, Y.~Song, L.~Shao, Nonconstant positive steady states and pattern
  formation of 1d prey-taxis systems, Journal of Nonlinear Science 22 (2017)
  71--97.

\bibitem{rev33}
K.~Wang, Q.~Wang, F.~Yu, Stationary and time-periodic patterns of two-predator
  and one-prey systems with prey-taxis, Discrete \& Continuous Dynamical
  Systems 37~(1) (2017) 505--543.

\bibitem{strings1}
P.~Avelino, D.~Bazeia, J.~Menezes, B.~de~Oliveira, String networks in
  lotka–volterra competition models, Physics Letters A 378~(4) (2014) 393 --
  397.

\bibitem{strings2}
P.~Avelino, D.~Bazeia, L.~Losano, J.~Menezes, B.~de~Oliveira, String networks
  with junctions in competition models, Physics Letters A 381~(11) (2017)
  1014--1020.

\end{thebibliography}

\end{document}